\documentclass[12pt,thmsa]{article}
\usepackage{amssymb}

\usepackage{sw20lart}

\input{tcilatex}
\begin{document}

\author{Krzysztof Ma\'{s}lanka\thanks{%
e-mail:maslanka@oa.uj.edu.pl} \\
Astronomical Observatory of the Jagiellonian University\\
ul. Orla 171\\
30-244 Cracow, POLAND}
\title{Hypergeometric-like Representation of the Zeta-Function of Riemann\thanks{%
Cracow Observatory preprint}}
\date{July 1997}
\maketitle

\begin{abstract}
We present a new expansion of the zeta-function of Riemann. It is given by
the formula (\ref{A1}) below.

The current formalism -- which combines both the idea of interpolation with
constraints and the concept of hypergeometric functions -- can, in a natural
way, be generalised within the theory of the zeta-function of Hawking
offering thus a variety of applications in quantum field theory, quantum
cosmology and statistical mechanics.\pagebreak
\end{abstract}

The aim of this paper is to prove a new expansion of the zeta-function of
Riemann (see e.g. \cite{Titch}) . Our approach combines both the idea
of interpolation with constraints and the theory of hypergeometric
functions. Further details can be found in \cite{Masl}.

\begin{theorem}
The following series: 
\begin{equation}
Z(s)=\frac{1}{s-1}\sum\limits_{k=0}^{\infty }\frac{\Gamma \left( k+1-\frac{s%
}{2}\right) }{\Gamma \left( 1-\frac{s}{2}\right) }\frac{A_{k}}{k!}
\label{A1}
\end{equation}
where 
\begin{equation}
A_{k}=\sum\limits_{j=0}^{k}(-1)^{j}\binom{k}{j}(2j+1)\zeta (2j+2)  \label{A2}
\end{equation}
are numerical coefficients, represents the zeta-function of Riemann 
\begin{equation}
\zeta (s)\stackrel{\text{df}}{=}\sum\limits_{m=1}^{\infty }m^{-s}  \label{A3}
\end{equation}
\end{theorem}

We call this expansion `hypergeometric-like' since, after
introducing an obvious symbolic notation, we may simply re-write (\ref{A1})
in a manifestly hypergeometric form 
\begin{equation}
Z(s)=\frac{1}{s-1}{}_{1}F_{0}\left( 1-\frac{s}{2};A\right) =\frac{1}{s-1}%
(1-A)^{\frac{s}{2}-1}  \label{A4}
\end{equation}
where successive $k$-th `powers' of $A$ are to be understood
as the appropriate coefficients $A_{k}$. In (\ref{A4}) we have made use of 
\begin{equation}
_{1}F_{0}(a;x)\equiv \sum\limits_{k=0}^{\infty }\frac{\Gamma \left(
k+a\right) }{\Gamma \left( a\right) }\frac{x^{k}}{k!}=\left( 1-x\right) ^{-a}
\label{A5}
\end{equation}

Contrary to the definition (\ref{A3}) in which, in order to assure
convergence, the condition $\func{Re}s>1$ must be satisfied, the series (\ref
{A1}) is convergent on the whole plane of the complex variable $s$, in
particular -- at the point $s=0$, which is important for various
applications in theoretical physics (see \cite{Hawking} for more details).
It should also be stressed out that, in spite of its appearance, the
definition (\ref{A2}) of coefficients $A_{k}$ is completely independent of
zeta itself since the latter, for positive and even arguments, may be
expressed with the use of the Bernoulli numbers\footnote{%
In spite of their appearence, formulae (\ref{A1}) and (\ref{A2}) are not
tautological. Indeed, by using 
\[
\zeta \left( 2n\right) =-\frac{\left( -1\right) ^{n}\left( 2\pi \right) ^{2n}%
}{2\left( 2n\right) !}B_{2n}\qquad (n=0,1,2,...) 
\]
where the Bernoulli numbers are 
\[
B_{n}=\sum\limits_{l=0}^{n}\frac{1}{l+1}\sum_{i=0}^{l}\left( -1\right) ^{i}%
\binom{l}{i}i^{n}\qquad (n=0,1,2,...) 
\]
the zeta function of Riemann, for positive and even arguments, may be
expressed using a finite number of simple arithmetic operations. Hence the
coefficients (\ref{A2}) may also be expressed in the form 
\[
A_{k}=\sum_{j=0}^{k}\binom{k}{j}\frac{\pi ^{2j+2}}{\left( 2\right)
_{j}\left( \frac{1}{2}\right) _{j}}B_{2j+2} 
\]
where $\left( a\right) _{k}$ denotes Pochhammer symbol. In other words,
according to (\ref{A1}) and (\ref{A2}), $\zeta (s)$ function of Riemann is
completely determined everywhere solely by its values in $s=2n,n=1,2,...,$
i. e. in the points where it can be computed independently (and exactly).},
i. e. with no explicit reference to the zeta-function itself.

\textbf{Proof}

Replacing in the definition (\ref{A2}) of $A_{k}$ the zeta function
according to its basic definition (\ref{A3}) and reversing the order of
summations gives 
\begin{eqnarray}
A_{k} &=&\sum\limits_{j=0}^{k}(-1)^{j}\binom{k}{j}(2j+1)\sum\limits_{n=1}^{%
\infty }\frac{1}{n^{2j+2}}=  \nonumber \\
&=&\sum\limits_{n=1}^{\infty }\frac{1}{n^{2}}\sum\limits_{j=0}^{k}(-1)^{j}%
\binom{k}{j}(2j+1)\left( \frac{1}{n}\right) ^{2j}  \label{B1}
\end{eqnarray}
Let us further note that 
\begin{equation}
(2j+1)\left( \frac{1}{n}\right) ^{2j}=n\stackunder{\alpha \rightarrow 1}{%
\lim }\frac{\text{d}}{\text{d}\alpha }\QOVERD( ) {\alpha }{n}^{2j+1}
\label{B2}
\end{equation}
where $\alpha $ is an auxiliary continuous parameter. Substituting (\ref{B2}%
) into (\ref{B1}) gives 
\begin{equation}
A_{k}=\stackunder{\alpha \rightarrow 1}{\lim }\frac{\text{d}}{\text{d}\alpha 
}\left[ \alpha \sum\limits_{n=1}^{\infty }\frac{1}{n^{2}}\sum%
\limits_{j=0}^{k}(-1)^{j}\binom{k}{j}\left( \frac{\alpha ^{2}}{n^{2}}\right)
^{j}\right]  \label{B3}
\end{equation}
The last sum in (\ref{B3}) may be recognized as the Newton binomial
expansion. Hence 
\begin{equation}
A_{k}=\stackunder{\alpha \rightarrow 1}{\lim }\frac{\text{d}}{\text{d}\alpha 
}\left[ \alpha \sum\limits_{n=1}^{\infty }\frac{1}{n^{2}}\left( 1-\frac{%
\alpha ^{2}}{n^{2}}\right) ^{k}\right]  \label{B4}
\end{equation}
giving thus yet another expression for the coefficients $A_{k}$.

Employing (\ref{B4}) in the expansion (\ref{A1}) yields, after some
rearrangements, 
\begin{eqnarray}
Z(s) &=&\frac{1}{s-1}\sum\limits_{k=0}^{\infty }\frac{\Gamma \left( k+1-%
\frac{s}{2}\right) }{\Gamma \left( 1-\frac{s}{2}\right) }\frac{A_{k}}{k!}=
\label{B5} \\
&=&\frac{1}{s-1}\stackunder{\alpha \rightarrow 1}{\lim }\frac{\text{d}}{%
\text{d}\alpha }\left[ \alpha \sum\limits_{n=1}^{\infty }\frac{1}{n^{2}}%
\sum\limits_{k=0}^{\infty }\frac{\Gamma \left( k+1-\frac{s}{2}\right) }{%
\Gamma \left( 1-\frac{s}{2}\right) }\frac{\left( 1-\frac{\alpha ^{2}}{n^{2}}%
\right) ^{k}}{k!}\right]   \nonumber
\end{eqnarray}
where the last sum coincides with the hypergeometric function $_{1}F_{0}$.
Thus 
\begin{eqnarray}
Z(s) &=&\frac{1}{s-1}\stackunder{\alpha \rightarrow 1}{\lim }\frac{\text{d}}{%
\text{d}\alpha }\left[ \alpha \sum\limits_{n=1}^{\infty }\frac{1}{n^{2}}{}%
_{1}F_{0}\left( 1-\frac{s}{2};1-\frac{\alpha ^{2}}{n^{2}}\right) \right] =
\label{B6} \\
&=&\frac{1}{s-1}\stackunder{\alpha \rightarrow 1}{\lim }\frac{\text{d}}{%
\text{d}\alpha }\alpha ^{1+s-2}\sum\limits_{n=1}^{\infty }\frac{1}{n^{2}}%
\left( \frac{1}{n^{2}}\right) ^{\frac{s}{2}-1}=  \nonumber \\
&=&\sum\limits_{n=1}^{\infty }\frac{1}{n^{s}}  \nonumber
\end{eqnarray}
which, according to the definition (\ref{A3}), is\footnote{%
More precisely, expansion (\ref{A1}) is an analytic continuation of the
series (\ref{A3}).} really $\zeta (s)$. In (\ref{B6}) we have made use of (%
\ref{A5}). This completes the proof of expansion (\ref{A1}).$\blacksquare $

\textbf{Discussion}

Coefficients (\ref{A2}) look suspiciously simple, even a bit tautological.
In order to understand how they actually work let us simply replace $\zeta
(2j+2)$ by unity (which is justifiable only for large $j$). Then we have 
\begin{equation}
a_{k}=\sum\limits_{j=0}^{k}(-1)^{j}\binom{k}{j}(2j+1)  \label{A21}
\end{equation}
and these are just $a_{0}=1,a_{1}=-2,a_{k}=0$ for $k=2,3,...$ which is
obviously convergent to zero. The following table contains integer numbers
which are under the sum (\ref{A21}) with $k$ labelling columns and $j$
labelling rows: 
\[
\begin{array}{ccccccc}
&  & {\tiny k=0} & {\tiny k=1} & {\tiny k=2} & {\tiny k=3} & {\tiny k=4} \\ 
&  &  &  &  &  &  \\ 
{\tiny j=0} &  & 1 & 1 & 1 & 1 & 1 \\ 
{\tiny j=1} &  & 0 & -3 & -6 & -9 & -12 \\ 
{\tiny j=2} &  & 0 & 0 & 5 & 15 & 30 \\ 
{\tiny j=3} &  & 0 & 0 & 0 & -7 & -28 \\ 
{\tiny j=4} &  & 0 & 0 & 0 & 0 & 9 \\ 
&  &  &  &  &  &  \\ 
a_{k} &  & 1 & -2 & 0 & 0 & 0
\end{array}
\]
The bottom row contains five initial $a_{k}$.The next table contains numbers
from the original sum (\ref{A2}), i.e. with $\zeta (2j+2)$ restored: 
\[
\begin{array}{ccccccc}
&  & {\tiny k=0} & {\tiny k=1} & {\tiny k=2} & {\tiny k=3} & {\tiny k=4} \\ 
&  &  &  &  &  &  \\ 
{\tiny j=0} &  & 1.\,645 & 1.\,645 & 1.\,645 & 1.\,645 & 1.\,645 \\ 
{\tiny j=1} &  & 0 & -3.25 & -6.49 & -9.74 & -12.99 \\ 
{\tiny j=2} &  & 0 & 0 & 5.09 & 15.26 & 30.52 \\ 
{\tiny j=3} &  & 0 & 0 & 0 & -7.03 & -28.11 \\ 
{\tiny j=4} &  & 0 & 0 & 0 & 0 & 9.01 \\ 
&  &  &  &  &  &  \\ 
A_{k} &  & 1.645 & -1.\,60 & 0.\,238 & 0.\,136 & 0.07\,21
\end{array}
\]
It is clear that upper rows differ significantly, especially the $0$-th row (%
$j=0$) which contains $\zeta (2)=\pi ^{2}/6=1.645...$ instead of $1$. As
before the bottom row contains five initial $A_{k}$.

Finally, it should be stressed out that the very fact of existence of
expansion (\ref{A1}) for the zeta of Riemann is by no means obvious. There
are many examples of regular, even elementary functions (e.g. the
exponential function) for which hypergeometric-like expansion do not
converge. Numerical experiments show that between equally spaced nodes there
are growing oscillations. The key feature here is that coefficients $A_{k}$
should tend to zero sufficiently fast which is not true for every function.

\end{document}